\documentclass[twocolumn]{svjour3}           
\pdfoutput=1   
\smartqed  
\usepackage{graphicx,psfrag,epsf,epsfig}
\usepackage{dcolumn}
\usepackage{bm,bbm}
\usepackage{pstricks,pst-plot}     
\usepackage{latexsym}
\usepackage{mathrsfs,amsmath,amssymb,textcomp}

 \journalname{Nanoscale Research Letters}
\begin{document}


\title{Theory and simulation of
photogeneration and transport in Si-SiO$_{x}$ superlattice absorbers}



\author{U. Aeberhard 
}


\institute{U. Aeberhard  \at
              IEF-5: Photovoltaik, Forschungszentrum J\"ulich, D-52425 J\"ulich,
Germany \\
             Tel: +49  2461 61 2615
             \\
              \email{u.aeberhard@fz-juelich.de} 
}

\date{Received: date / Accepted: date}

\maketitle

\begin{abstract}
Si-SiO$_{x}$ superlattices are among the candidates that have been proposed as high 
band gap absorber material in all-Si tandem solar cell devices. Due to the large potential
barriers for photoexited charge carriers, transport in these devices is restricted to quantum confined superlattice states. 
As a consequence of the finite number of wells, large built-in fields and any kind of disorder, 
the electronic spectrum can deviate considerably from the minibands of a regular superlattice. In this paper, a
quantum-kinetic theory based on the non-equilibrium Green's function formalism
for an effective mass Hamiltonian is used to investigate photogeneration and
transport in such devices for arbitrary geometry and operating conditions. By including the coupling of electrons to
both photons and phonons, the theory is able to provide a microscopic picture of indirect generation, 
carrier relaxation and inter-well transport mechanisms beyond the ballistic regime.
\end{abstract}

\section{\label{sec:intro}Introduction} 
Si-SiO$_{x}$ superlattices have been proposed as candidates for the high band
gap absorber component in all-Si tandem solar cells \cite{green:00,green:01}.
In these devices, photocurrent flow is enabled via the overlap of states in neighboring 
Si quantum wells separated by ultra-thin
oxide layers, i.e. unlike in the case of an intermediate band solar
cell, the superlattice states contribute to the
optical transitions and at the same time provide transport of photocarriers,
which makes it necessary to control both the optical and the transport
properties of the multilayer structure. To this end, a suitable theoretical
picture of the optoelectronic processes in such type of structures is highly desirable.

There are several peculiar aspects of the device which require special
consideration in the choice of an appropriate model. First of all, a microscopic
model for the electronic structure is indispensable, since the relevant states
are those of an array of strongly coupled quantum wells. In a standard approach, 
these states are described with simple Kronig-Penney models for a regular,
infinitely extended superlattice. The superlattice dispersion obtained in this way can then be used
to determine an effective density of states as well as the absorption
coefficient to be used in macroscopic 1D solar cell device simulators. However, 
depending on the internal field and the structural disorder, the
heterostructure states may deviate considerably from regular minibands or
can even form Wannier-Stark ladders. Furthermore, the charge carrier mobility,
which has a crucial impact on the charge collection efficiency in solar cells,
depends on the dominant transport regime at given operating
conditions, which may be described by miniband transport, sequential tunneling
or Wannier-Stark hopping \cite{wacker:02}, relying on processes
that are not accessible to standard macroscopic transport models. 
 
In this paper, the photovoltaic properties of quantum well superlattice
absorbers are investigated numerically on the example of a Si-SiO$_{x}$ multilayer 
structure embedded in the intrinsic region of a $p$-$i$-$n$ diode, using a
multiband effective mass approximation for the electronic structure and the 
non-equilibrium Green's function (NEGF) formalism for
inelastic quantum transport, which permits to treat on equal footing both
coherent and incoherent transport as well as phonon-assisted optical
transitions at arbitrary internal fields and heterostructure potentials.

\section{\label{sec:model}Theoretical Model}
In order to enable a sound theoretical description of the pivotal photovoltaic
processes in semiconductor nanostructures, i.e. charge carrier generation,
recombination and collection, both optical transitions and inelastic quantum
transport are to be treated on equal footing within a consistent microscopic
model. To this end, a theoretical framework based on the NEGF formalism was
developed \cite{ae:prb_08,ae:thesis} and applied to quantum well solar cell
devices. Here, we reformulate the theory for a multiband effective mass
Hamiltonian, similar to \cite{steiger:iwce_09,steiger:thesis}, and extend it 
to cover the phonon-assisted indirect transitions that dominate the photovoltaic
processes in Si-based devices. Furthermore, in difference to the former case,
both photogeneration and transport processes take place within superlattice
states, since escape of carriers to continuum states is not possible due to the
large band offsets. 
 
\subsection{Hamiltonian and basis}
The full quantum photovoltaic
device  is described in terms of the model Hamiltonian
\begin{align}
 \hat{H}=&\hat{H}_{e}+\hat{H}_{\gamma}+\hat{H}_{p},&\mathrm{(total)}\qquad
 \\ \hat{H}_{e}=&\hat{H}_{e}^{0}+\hat{H}_{e}^{i},
&\mathrm{(electronic)}\qquad \\ 
\hat{H}_{e}^{i}=&\hat{H}_{e\gamma}+\hat{H}_{ep}+\hat{H}_{ee},
&\mathrm{(interaction)}\qquad
\end{align}
consisting of the coupled systems of electrons ($\hat{H}_{e}$), photons
($\hat{H}_{\gamma}$) and phonons ($\hat{H}_{p}$). Since the focus is on
the electronic device characteristics, only $\hat{H}_{e}$ is considered here,
however including all of the terms corresponding to coupling to the bosonic systems. 

The electronic system without coupling to the bosonic degrees of freedom is described by 
\begin{align}
\hat{H}_{e}^{0}=&-\frac{\hbar^{2}}{2m_{0}}\Delta+\tilde{U}(z),
\end{align}
with
\begin{align}
\tilde{U}(z)=U(z)+V_{0}(z),
\end{align}
where $V_{0}$ is the heterostructure potential and $U$ is the Hartree term of the Coulomb interaction corresponding
to the solution of Poisson's equation that considers carrier-carrier interactions ($\hat{H}_{ee}$) on a mean-field level.

The Hamiltonian representations for the interaction terms are obtained starting from the single particle 
interaction potentials. For the electron-photon interaction, the latter is given via the linear coupling to
the vector potential operator of the electromagnetic field  $\hat{{\mathbf A}}$,
\begin{align}
 \hat{H}_{e\gamma}=-\frac{e}{m_{0}}\hat{{\mathbf A}}\cdot\hat{{\mathbf p}}
\end{align}
with $\hat{{\mathbf p}}$ the momentum operator and 
\begin{align}
\hat{\mathbf{A}}({\mathbf r},t)=&\sum_{\lambda,{\mathbf
q}}\left[\mathbf{A}_{0}(\lambda,\mathbf{q}) \hat{b}_{\lambda,{\mathbf
q}}(t)+\mathbf{A}_{0}^{*}(\lambda,-\mathbf{q}) 
\hat{b}_{\lambda,{-\mathbf q}}^{\dagger}(t)\right]\nonumber\\
&\times e^{i{\mathbf q}{\mathbf
r}},\label{eq:photfieldop}\\
\mathbf{A}_{0}(\lambda,\mathbf{q})=&\frac{\hbar}{\sqrt{2\epsilon_{0}V\hbar\omega_{\mathbf{q}}}}
\boldsymbol{\epsilon}_{\lambda{\mathbf q}},
\end{align}
where ${\mathbf \epsilon}_{\lambda{\mathbf q}} $ is the polarization of the
photon with wave vector 
${\mathbf q}$ and energy $\hbar\omega_{\mathbf{q}}$ added to or removed from photon
mode $(\lambda,\mathbf{q})$ by the bosonic creation and annihilation operators 
\begin{align}
\hat{b}_{\lambda,{\mathbf q}}^{\dagger}(t)=&\hat{b}_{\lambda,{\mathbf
q}}^{\dagger}e^{i\omega_{{\mathbf q}}t},\quad \hat{b}_{\lambda,{\mathbf
q}}(t)=\hat{b}_{\lambda,{\mathbf q}}e^{-i\omega_{{\mathbf q}}t},
\end{align}
and $V$ is the absorbing volume. 

The vibrational degrees of freedom of the system are described in terms of the coupling of the force field 
of the electron-ion potential $V_{ei}$ to the 
quantized field $\boldsymbol{\mathcal{\hat{U}}}$ of the ionic displacement \cite{schaefer:02}, 
\begin{align}
\hat{H}_{ep}({\mathbf
r},t)=&\sum_{\mathbf{L},\boldsymbol{\kappa}}\boldsymbol{\mathcal{\hat{U}}}(\mathbf{L}+\boldsymbol{\kappa},t)\cdot
\nabla V_{ei}[\mathbf{r}-(\mathbf{L}+\boldsymbol{\kappa})],
\end{align}
with the displacement field given by the Fourier expansion
\begin{align}
\mathcal{\hat{U}}_{\alpha}(\mathbf{L}\boldsymbol{\kappa},t)=&\sum_{\Lambda,\mathbf{Q}}
\mathcal{U}_{\alpha\boldsymbol{\kappa}}(\Lambda,\mathbf{Q})
e^{i\mathbf{Q}\cdot(\mathbf{L}+\boldsymbol{\kappa})}
\big[\hat{a}_{\Lambda,\mathbf{Q}}(t)+\hat{a}^{\dagger}_{\Lambda,-\mathbf{Q}}(t)\big], 
\end{align}
where the ion equilibrium position is $\mathbf{L}+\boldsymbol{\kappa}$,
with $\mathbf{L}$ the lattice position and $\boldsymbol{\kappa}$ the relative
position of a specific basis atom at this lattice site, and
$\hat{a}_{\Lambda,{\mathbf Q}},\hat{a}_{\Lambda,{\mathbf Q}}^{\dagger}$ are 
the bosonic creation and annihilation operators for a (bulk) phonon mode with polarization
$\Lambda$ and wave vector ${\mathbf Q}$ in the first Brillouin zone. The
corresponding polarization vector is $\boldsymbol{\epsilon}
_{\alpha\mathbf{\kappa}\Lambda}(\mathbf{Q})$. 
 
For numerical implementation of the model, the above Hamiltonian needs to be
represented in a suitable basis. Due to the amorphous nature of the SiO$_{x}$
layers, atomistic models are of limited applicability. Furthermore, the use of an effective mass theory
 simplifies the electronic model considerably. For a quasi-onedimensional
multilayer system, where quantization appears only in the vertical (growth) direction, 
the corresponding basis functions have the form
\begin{align}
\psi_{in\mathbf{k}_{\parallel}}(\mathbf{r})=\varphi_{i\mathbf{k}_{\parallel}}(\mathbf{r})
u_{n\mathbf{k}_{0}}(\mathbf{r}),\label{eq:basis}
\end{align}
where $\varphi_{i\mathbf{k}_{\parallel}}$ is the envelope basis function for
discrete spatial (layer) index $i$ (longitudinal) and transverse momentum
$\mathbf{k}_{\parallel}$,   $u_{n\mathbf{k}_{0}}$ is the  Bloch function of
bulk band $n$, centered on $\mathbf{k}_{0}$. In the case of a system with
large transverse extension, the envelope basis function can be written as
\begin{align}
\varphi_{i\mathbf{k}_{\parallel}}(\mathbf{r})=\frac{e^{i\mathbf{k}_{\parallel}\mathbf{r}_{\parallel}}}{\sqrt{S}}\chi_{i}(z),
\end{align}
where $\mathbf{r}_{\parallel}=(x,y)$, $S$ is the cross sectional area and
$\chi_{i}$ is the localized longitudinal envelope function basis element. For the latter, finite element shape 
functions are a popular
choice \cite{steiger:thesis,kubis:09}. Here, we will use a simple finite
difference basis equivalent to a separate single band tight-binding approach
for each band \cite{lake:97,henrickson:02,jin:06}.
In the above basis, the fermion field operators for the charge carriers are
represented via
\begin{align}
\hat{\Psi}(\mathbf{r},t)&=\sum_{i,n,\mathbf{k}_{\parallel}}\psi_{in\mathbf{k}_{\parallel}}(\mathbf{r})
\hat{c}_{in\mathbf{k}_{\parallel}}(t),\label{eq:fieldop_1}\\
\hat{\Psi}^{\dagger}(\mathbf{r},t)&=\sum_{i,n,\mathbf{k}_{\parallel}}\psi^{*}_{in\mathbf{k}_{\parallel}}(\mathbf{r})
\hat{c}^{\dagger}_{in\mathbf{k}_{\parallel}}(t),\label{eq:fieldop_2}
\end{align}
where $\hat{c}^{\dagger},\hat{c}$ are single fermion creation and annihilation
operators. The representation of the model system Hamiltonian in the above basis
is now obtained in standard second quantization, i.e.
\begin{align}
\mathcal{H}(t)&=\int d^{3}r 
\hat{\Psi}^{\dagger}(\mathbf{r},t)\hat{H}_{e}^{0}\hat{\Psi}(\mathbf{r},t)\\
&=\sum_{i,j}\sum_{n,m}\sum_{\mathbf{k}_{\parallel}}H_{in;jm}(\mathbf{k}_{\parallel})
\hat{c}^{\dagger}_{in\mathbf{k}_{\parallel}}(t)
\hat{c}_{jm\mathbf{k}_{\parallel}}(t).
\end{align}

 \subsection{Green's functions, self energies and quantum kinetic equations}
 Within the non-equilibrium Green's function theory of quantum optics and transport in excited 
semiconductor nanostructures, physical quantities are expressed in terms of quantum statistical 
ensemble averages of single particle operators for the interacting
quasiparticles introduced above, namely the fermion field operator $\hat{\Psi}$ for the charge
carriers, the quantized photon field vector potential $\hat{\mathbf{A}}$ for the
photons and the ionic displacement field $\hat{\boldsymbol{\mathcal{U}}}$ for the
phonons. The corresponding Green's functions are
\begin{align}
 \mathcal{G}(\underbar{1},\underbar{2})=&-\frac{i}{\hbar}\langle\hat{\Psi}(\underbar{1})
 \hat{\Psi}^{\dagger}
 (\underbar{2})\rangle_{\mathcal{C}},\qquad&\mathrm{(electrons)}\\
 \mathcal{D}_{ik}^{\gamma}(\underbar{1},\underbar{2})=&-\frac{i}{\hbar}\langle \hat{A}_{i}(\underbar{1})
 \hat{A}_{k}(\underbar{2})\rangle_{\mathcal{C}},\qquad&\mathrm{(photons)}\\
  \mathcal{D}_{\alpha\beta}^{p}(\underbar{1},\underbar{2})=&-\frac{i}{\hbar}
  \langle\hat{\mathcal{U}}_{\alpha}
  (\underbar{1})\hat{\mathcal{U}}_{\beta}(\underbar{2})
  \rangle_{\mathcal{C}},\qquad&\mathrm{(phonons)}  
 \end{align}   
where $\langle...\rangle_{C}$ denotes the contour ordered operator average 
peculiar to non-equilibrium quantum statistical mechanics
\cite{kadanoff:62,keldysh:65} for arguments
$\underbar{1}=(\mathbf{r}_{1},t_{1})$ with temporal components on the Keldysh
contour\cite{keldysh:65}.

The Green's functions follow as the solutions to corresponding \emph{Dyson's
equations} \cite{henneberger:88_3,pereira:96,pereira:98,schaefer:02},
\begin{align}
\int
d\underbar{3}\left[\mathcal{G}_{0}^{-1}(\underbar{1},\underbar{3})-\Sigma(\underbar{1},\underbar{3})\right]
\mathcal{G}(\underbar{3},\underbar{2}) &=\delta(\underbar{1}-\underbar{2}),\nonumber\\
\int d\underbar{3}
\left[(\overleftrightarrow{\boldsymbol{\mathcal{D}}}_{0}^{\gamma})^{-1}(\underbar{1},\underbar{3})
- \overleftrightarrow{\boldsymbol{\Pi}}^{\gamma}(\underbar{1},\underbar{3})\right]
 \overleftrightarrow{\boldsymbol{\mathcal{D}}}^{\gamma}(\underbar{3},\underbar{2})&=
 \overleftrightarrow{\boldsymbol{\delta}}(\underbar{1}-\underbar{2}),\nonumber\\
\int d\underbar{3}
\left[(\boldsymbol{\mathcal{D}}_{0}^{p})^{-1}(\underbar{1},\underbar{3})-
\boldsymbol{\Pi}^{p}(\underbar{1},\underbar{3})\right]
\boldsymbol{\mathcal{D}}^{p}(\underbar{3},\underbar{2})&=
\boldsymbol{\delta}(\underbar{1}-\underbar{2}).
 \end{align}
$\mathcal{G}_{0}$, $\mathcal{D}_{0}^{\gamma}$ and $\mathcal{D}_{0}^{p}$ are the
propagators for noninteracting electrons, photons and phonons, respectively,  $\leftrightarrow$ denotes
 transverse and boldface tensorial quantities. The
 electronic self-energy $\Sigma$ encodes the renormalization of the charge carrier Green's
functions due to the interactions with photons and phonons, i.e. generation, recombination and
 relaxation processes. Charge injection and absorption at 
contacts is considered via an additional boundary self-energy term reflecting the openness of 
the system. The photon and phonon self-energy tensors $\overleftrightarrow{\boldsymbol{\Pi}}^{\gamma}$ and
$\boldsymbol{\Pi}^{p}$ describe the renormalization of the optical and vibrational
modes, leading to phenomena such as photon recycling or the phonon bottleneck 
responsible for hot carrier effects. The self-energies can be derived either via
perturbative methods using a diagrammatic approach or a Wick factorization or using variational
derivatives. In the following, any renormalizing effect of
the electronic system on the photons and phonons is neglected, i.e. the coupling to the bosons
corresponds to the connection to corresponding equilibrium reservoirs. While
this treatment is generally a good approximation in the case of phonons, it
is valid for the coupling to the photonic systems only in the case of low
absorption, i.e. weak coupling or very short absorber length. 

The use of the equilibrium boson propagators implies 
that only the electronic
Dyson equations are solved. In the chosen discrete real-space basis, the components of the steady-state Dyson and
Keldysh equations for electronic Greens functions are turned into a linear system\footnote{In steady-state, the
Green's functions depend only on the difference $\tau=t-t'$ of the real-time variables, which
is Fourier-transformed to energy.}($\nu=\mathbf{k}_{\parallel},E$) 
\begin{align}
\mathbf{G}^{R}(\nu)&=\left[\big\{\mathbf{G}_{0}^{R}(\nu)\big\}^{-1}
-\mathbf{\Sigma}^{R I}(\nu)-
\mathbf{\Sigma}^{RB}(\nu)\right]^{-1},\label{eq:retgf}\\
\mathbf{G}_{0}^{R}(\nu)&=\left[(E+i\eta)\mathbbm{1}-\mathbf{h}(\mathbf{k}_{\parallel})\right]^{-1},\\
\mathbf{G}^{A}(\nu)&=[\mathbf{G}^{R}(\nu)]^{\dagger},\\
\mathbf{G}^{\lessgtr}(\nu)&=\mathbf{G}^{R}(\nu)
\left[\mathbf{\Sigma}^{\lessgtr I}(\nu)
+\mathbf{\Sigma}^{\lessgtr
B}(\nu)\right]\mathbf{G}^{A}(\nu),\label{eq:corrf}
\end{align}
for each total energy $E$ and transverse momentum $\mathbf{k}_{\parallel}$.
There are two types of self-energies in the
above equations.  The term $\Sigma^{\cdot I}$ is due to the interactions between
the different degrees of freedom of the system. The expressions for electron-photon and
electron-phonon interaction are determined as the Fock term within many-body
perturbation theory on the level of a selfconsistent Born approximation, and 
using the equilibrium boson propagators, they are obtained in the following form, 
($\alpha=\gamma,p$)
\begin{align}
\boldsymbol{\Sigma}_{e\alpha}^{\lessgtr}({\mathbf k}_{\parallel};E)=&
\sum_{\lambda,\mathbf{q}}\boldsymbol{\mathcal{M}}^{e\alpha}({\mathbf
k}_{\parallel},\mathbf{q},\lambda)
\big[N^{\alpha}_{\lambda,\mathbf{q}}\mathbf{G}^{\lessgtr}({\mathbf
k}_{\parallel};E\mp\hbar\omega_{\lambda,\mathbf{q}})\nonumber\\
&+(N^{\alpha}_{\lambda,\mathbf{q}}
+1)\mathbf{G}^{\lessgtr} ({\mathbf
k}_{\parallel};E\pm\hbar\omega_{\lambda,\mathbf{q}})\big]\nonumber\\
&\times\boldsymbol{\mathcal{M}}^{e\alpha}({\mathbf
k}_{\parallel},-\mathbf{q},\lambda)\label{eq:intse_lg}
\end{align}
and 
\begin{align}
\mathbf{\Sigma}_{e\alpha}^{R,A}({\mathbf
k}_{\parallel};E)&=i\int\frac{dE'}{2\pi}\frac{\boldsymbol{\Sigma}_{e\alpha}^{>}(,\mathbf{k}_{\parallel};E')
-\boldsymbol{\Sigma}_{e\alpha}^{<}(\mathbf{k}_{\parallel};E')}{E'-E\pm
i\eta}\nonumber\\
&=\mathcal{P}\int\frac{dE'}{2\pi}
\frac{\boldsymbol{\Gamma}(\mathbf{k}_{\parallel};E')}{E'-E}
\mp\frac{i}{2}\boldsymbol{\Gamma}(\mathbf{k}_{\parallel};E),
\label{eq:intse_ra}
\end{align}
where
\begin{align}
\boldsymbol{\Gamma}(\mathbf{k}_{\parallel};E)=i\left[\boldsymbol{\Sigma}_{e\alpha}^{>}(\mathbf{k}_{\parallel};E)
-\boldsymbol{\Sigma}_{e\alpha}^{<}(\mathbf{k}_{\parallel};E)\right].
\end{align}
Since the principal value integral corresponds to the real part of the self
energy and thus to the renormalization of the electronic structure, which is
both small and irrelevant for the photovoltaic performance, it is neglected in
the numerical implementation.

Once the Green's functions and self-energies have been determined via self-consistent solution of Eqs. 
 \eqref{eq:retgf}-\eqref{eq:corrf} and \eqref{eq:intse_lg}-\eqref{eq:intse_ra}, they can be directly used to express the
 physical quantities that characterize the system, such as charge carrier and current densities as well as the rates for the 
 different scattering processes.
 
 \subsection{Microscopic optoelectronic conservation laws and scattering rates}
 The macroscopic balance equation for a photovoltaic system is
the steady state continuity equation for the charge carrier density
\begin{align}
\nabla\cdot 
\mathbf{j}_{c}(\mathbf{r})=\mathcal{G}_{c}(\mathbf{r})-\mathcal{R}_{c}(\mathbf{r}),\quad
c=e,h, 
\label{eq:macro_cont}
\end{align}
where $j_{c}$ is the particle current density, $\mathcal{G}_{c}$ the generation rate and $\mathcal{R}_{c}$ the
recombination rate of carriers species $c$\footnote{The dimensions are that of
a volume rate, $[\mathcal{G},\mathcal{R}]=m^{-3}s^{-1}$.}.
In the microscopic theory, the divergence of the electron (particle) current is given by \cite{kadanoff:62,keldysh:65}
\begin{align}
\nabla\cdot
\mathbf{j}(\mathbf{r})
=-&\frac{2}{V}\int\frac{dE}{2\pi\hbar}\int d^3 r'\Big[\Sigma^{R}
(\mathbf{r},\mathbf{r}';E)G^{<}(\mathbf{r}',\mathbf{r};E)
\nonumber\\+&\Sigma^{<}(\mathbf{r},\mathbf{r}';E)G^{A}(\mathbf{r}',\mathbf{r};E)
-G^{R}(\mathbf{r},\mathbf{r}';E)\nonumber\\&\times\Sigma^{<}(\mathbf{r}',\mathbf{r};E)-
G^{<}(\mathbf{r},\mathbf{r}';E)\Sigma^{A}(\mathbf{r}',\mathbf{r};E)\Big].\label{eq:currcons}
\end{align}
If the integration is restricted to either conduction or valence bands,
the above equation corresponds to the microscopic version of \eqref{eq:macro_cont}
and provides on the RHS  the total \emph{local} interband scattering rate. The total interband 
current is found by integrating the
divergence over the absorbing/emitting volume, and is equivalent to the total
\emph{global} transition rate and, via the Gauss theorem, to the
difference of the interband currents at the boundaries of the interacting
region. Making use of the cyclic property of the trace, it can be
expressed in the form
\begin{align}
R=&\frac{2}{V}\int d^3 r\int\frac{dE}{2\pi\hbar}\int d^3
r'\Big[\Sigma^{<}(\mathbf{r},\mathbf{r}';E)G^{>}(\mathbf{r}',\mathbf{r};E)\nonumber\\
&-\Sigma^{>}(\mathbf{r},\mathbf{r}';E)G^{<}(\mathbf{r}',\mathbf{r};E)
\Big],\label{eq:totrate}
\end{align}
with units $[R]=s^{-1}$. If we are interested in the interband scattering rate, we can
neglect in Eq. \eqref{eq:totrate} the contributions to the
self-energy from \emph{intraband} scattering, e.g. via interaction with phonons,
low energy photons (free carrier absorption) or ionized impurities, since
they cancel upon energy integration over the band. Since inequivalent
conduction band valleys may be described by different bands, the corresponding 
inter-valley scattering process has also interband character with a nonvanishing rate, 
as long as only one of the valleys is considered in the rate evaluation. 
Furthermore, if self-energies and Green's functions are determined
self-consistently as they must in order to guarantee current conservation, 
the Green's functions are related to the scattering self-energies via the Dyson
equation for the propagator and the Keldysh equation for the correlation functions as given in Eqs.
 \eqref{eq:retgf}-\eqref{eq:corrf}, and will thus be modified due to the
 intraband scattering. In the present case of indirect optical transitions, the
 Greens functions entering the rate for electron-photon scattering between the $\Gamma$ bands are the solutions of
 Dyson equations with an intervalley phonon scattering self-energy and may thus contain
 contributions from the $X$-valleys. In the same way, the $\Gamma_{c}$ Greens
 functions entering the electron-phonon $\Gamma_{c}-X$ scattering rate contain
 a photogenerated contribution. By this way, indirect,
 phonon-assisted optical transitions are enabled.

\section{\label{sec:numerics}Implementation for Si-SiO$_{x}$
superlattice absorbers}

\subsection{Electronic structure model}
Within the EMA for silicon chosen for this work, the electrons are described by
a multi-valley picture with different values for transverse and longitudinal
effective mass, similar to \cite{jin:06_2}. However, for
simplicity, in the case of transverse $X$ valleys ($X_{\parallel}$), the anisotropy 
in the transverse mass is neglected and an average value is used. The virtual $\Gamma$ states used in the
indirect transitions are described by an additional (negative) mass. The holes
are modelled by two decoupled single bands with different effective masses
corresponding to heavy and light holes. Thus, in total, five bands are used to
describe the electronic structure, three for the electrons
($X_{\parallel},~X_{\perp},~\Gamma_{c}$) and two for the holes
($\Gamma_{vl},~\Gamma_{vh}$). The band parameters used in the simulations are
listed in Tab. \ref{tab:band_parameters}. 
For each band, a set of Green's functions are computed from the corresponding
decoupled Dyson and Keldysh equations. In the computation of physical
quantities such as electron and hole densities as well as the corresponding
current densities, the summation over all conduction or valence bands needs to be
performed, 
\begin{align}
n_{i}=&\sum_{b=\Gamma_c,X_{\parallel},X_{\perp}}f_{b}n_{i,b}\\
=&\sum_{b}f_{b}\sum_{\mathbf{k}_{\parallel}}\int\frac{dE}{\pi
A\Delta}(-i)G^{<}_{ii,b}(\mathbf{k}_{\parallel};E),
\end{align}
where $f_{b}$ denotes the degeneracy of the conduction bands, which is 
$f_{\Gamma_c}=1$, $f_{X_{\parallel}}=4$ and $f_{X_{\perp}}=2$. Similarly, the
electron current in terms of the Green's functions reads
\begin{align}
J_{i}=&\sum_{b=\Gamma_c,X_{\parallel},X_{\perp}}f_{b}J_{i,b}\\
=&\sum_{b}f_{b}\sum_{\mathbf{k}_{\parallel}}\int\frac{dE}{\pi\hbar
A}\big[t_{ii+1}G^{<}_{i+1i,b}(\mathbf{k}_{\parallel};E)\nonumber\\&-t_{i+1i}G^{<}_{ii+1,b}(\mathbf{k}_{\parallel};E)\big].
\end{align}
For the chosen model of the bulk band structure, the total radiative rate is
\begin{align}
R_{e\gamma}
=&\frac{2}{\hbar\mathcal{A}}\int_{\Gamma_{c}}\frac{dE}{2\pi}\mathrm{Tr}\Big\{\sum_{\mathbf{k}_{\parallel}}
\big[\mathbf{\Sigma}_{e\gamma,\Gamma_{c}}^{>}(\mathbf{k}_{\parallel};E)\mathbf{G}_{\Gamma_{c}}^{<}(\mathbf{k}_{\parallel};E)
\nonumber\\
&-\mathbf{\Sigma}_{e\gamma,\Gamma_{c}}^{<}(\mathbf{k}_{\parallel};E)\mathbf{G}_{\Gamma_{c}}^{>}(\mathbf{k}_{\parallel};E)
\big]\Big\},\label{eq:totratelayer}
\end{align}
and the inter-valley phonon scattering rate reads
\begin{align}
&R_{ep,\Gamma-X}
=\frac{2}{\hbar\mathcal{A}}\int_{\Gamma_{c}}\frac{dE}{2\pi}\mathrm{Tr}\Big\{\sum_{\mathbf{k}_{\parallel}}
\big[\mathbf{\Sigma}_{ep(\Gamma-X),\Gamma_{c}}^{>}(\mathbf{k}_{\parallel};E)\nonumber\\
&\times\mathbf{G}_{\Gamma_{c}}^{<}(\mathbf{k}_{\parallel};E)
-\mathbf{\Sigma}_{ep(\Gamma-X),\Gamma_{c}}^{<}(\mathbf{k}_{\parallel};E)\mathbf{G}_{\Gamma_{c}}^{>}(\mathbf{k}_{\parallel};E)
\big]\Big\}.
\end{align}

\begin{table}[b]
\begin{center}
\caption{\label{tab:band_parameters} Band parameters used in simulations}
\begin{tabular}{lcc}
\hline\noalign{\smallskip}
&Si &SiO$_{x}$ \\
\noalign{\smallskip}\hline\noalign{\smallskip}
 $m^{*}_{\Gamma c}/m_{0}$&  -0.3 & -0.3 \\
 $m^{*}_{X_{\parallel}}/m_{0}$& 0.98& 0.4\\
 $m^{*}_{X_{\perp}}/m_{0}$&0.19& 0.4\\
$m^{*}_{\Gamma v,lh}/m_{0}$&0.16& 0.4\\ 
$m^{*}_{\Gamma v,hh}/m_{0}$&0.49 & 0.4\\
$E_{g,\Gamma v-\Gamma c}$ [eV]& 3.5& 5.5\\
$E_{g,\Gamma v-X}$ [eV]& 1.1&3.1\\
\noalign{\smallskip}\hline
\end{tabular}
\end{center}
\end{table}

\subsection{Interaction parameters}

Optical transitions are assumed to take place only at the center of the
Brillouin zone, i.e. between $\Gamma_v$ and virtual $\Gamma_c$ states, the
latter being (de-)populated via phonon scattering from (to) the $X$ valleys, which carry the
photocurrent. All other transition channels, e.g. phonon scattering in the
valence band prior to photon absorption, are neglected at this stage. The
momentum matrix element in the electron-photon coupling is thus to be taken
between the $\Gamma_v$ and $\Gamma_c$ bands at $\mathbf{k}_{0}=\mathbf{0}$. The
interaction matrix elements are evaluated using an average effective coupling for both light and heavy holes.

Four different types of phonons are used in the present work to describe both
carrier relaxation as well as phonon assisted optical transitions. For the
relaxation process, $X-X$ inter-valley scattering is used for the electrons and
non-polar optical phonon scattering for the holes. Further broadening is added
for both carrier species through acoustic phonon scattering in the deformation
potential formulation. Finally, the momentum transfer for the indirect optical
transitions is mediated via $\Gamma_{c}-X$ intervalley scattering.

\section{\label{sec:results} Numerical results and discussion}

\subsection{Model system}
\begin{figure*}[t!]
\begin{minipage}{9cm}
\begin{center}
 \includegraphics[height=6cm]{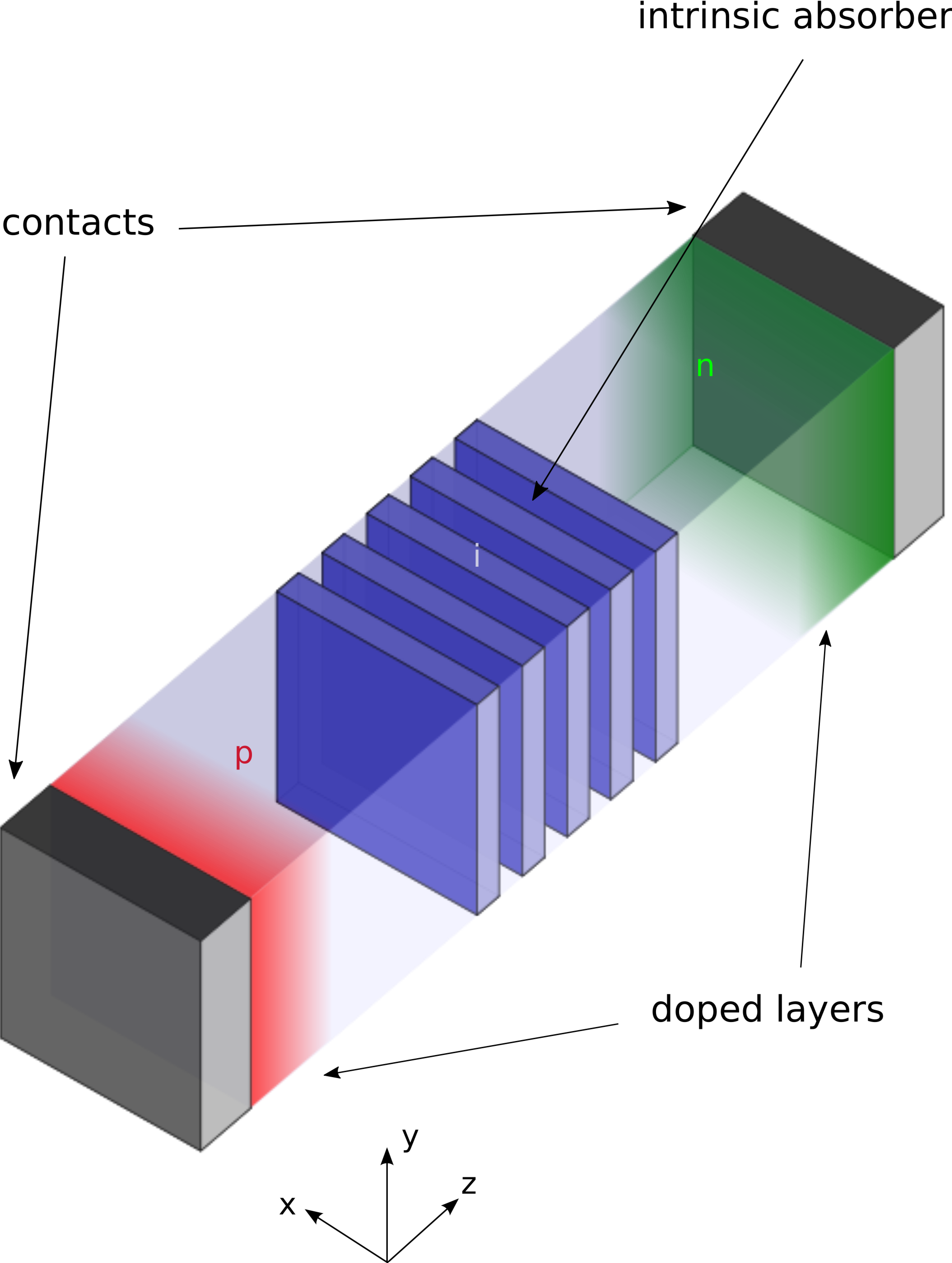}
 \end{center}
\caption{Spatial structure of the $p$-$i(SL)$-$n$ model system.\label{fig:struct}} 
\end{minipage}
\begin{minipage}{9cm}
\begin{center} 
\includegraphics[height=5.5cm]{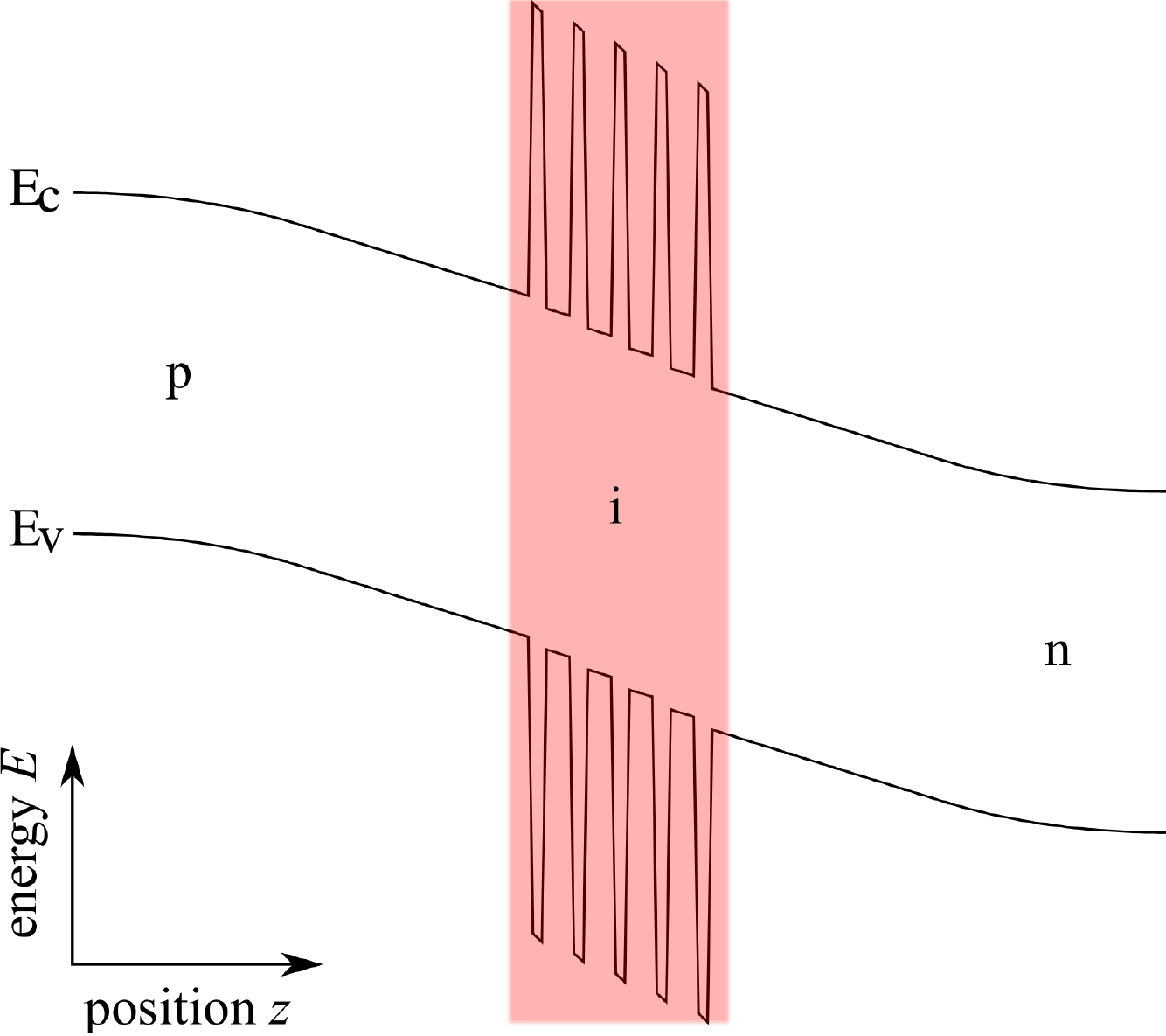}
 \end{center}
\vspace{5mm}
\caption{Band diagram of the $p$-$i(SL)$-$n$ model system.\label{fig:banddiag}}
\end{minipage}
\end{figure*}
The model system under investigation is shown schematically in Fig. \ref{fig:struct}. It
consists of a set of four coupled quantum wells of 6 monolayer (ML) width with layers separated by oxide barriers of 3 ML
thickness, embedded in the intrinsic region of a Si $p$-$i$-$n$ diode. The thickness of the 
doped layers is 50 ML, while the total length of the $i$-region amounts to 154 ML. 
This composition and doping leads to the band diagram shown in
Fig. \ref{fig:banddiag}.

\subsection{Density of states}
\begin{figure}[b!]
\begin{center}
 \includegraphics[height=6cm]{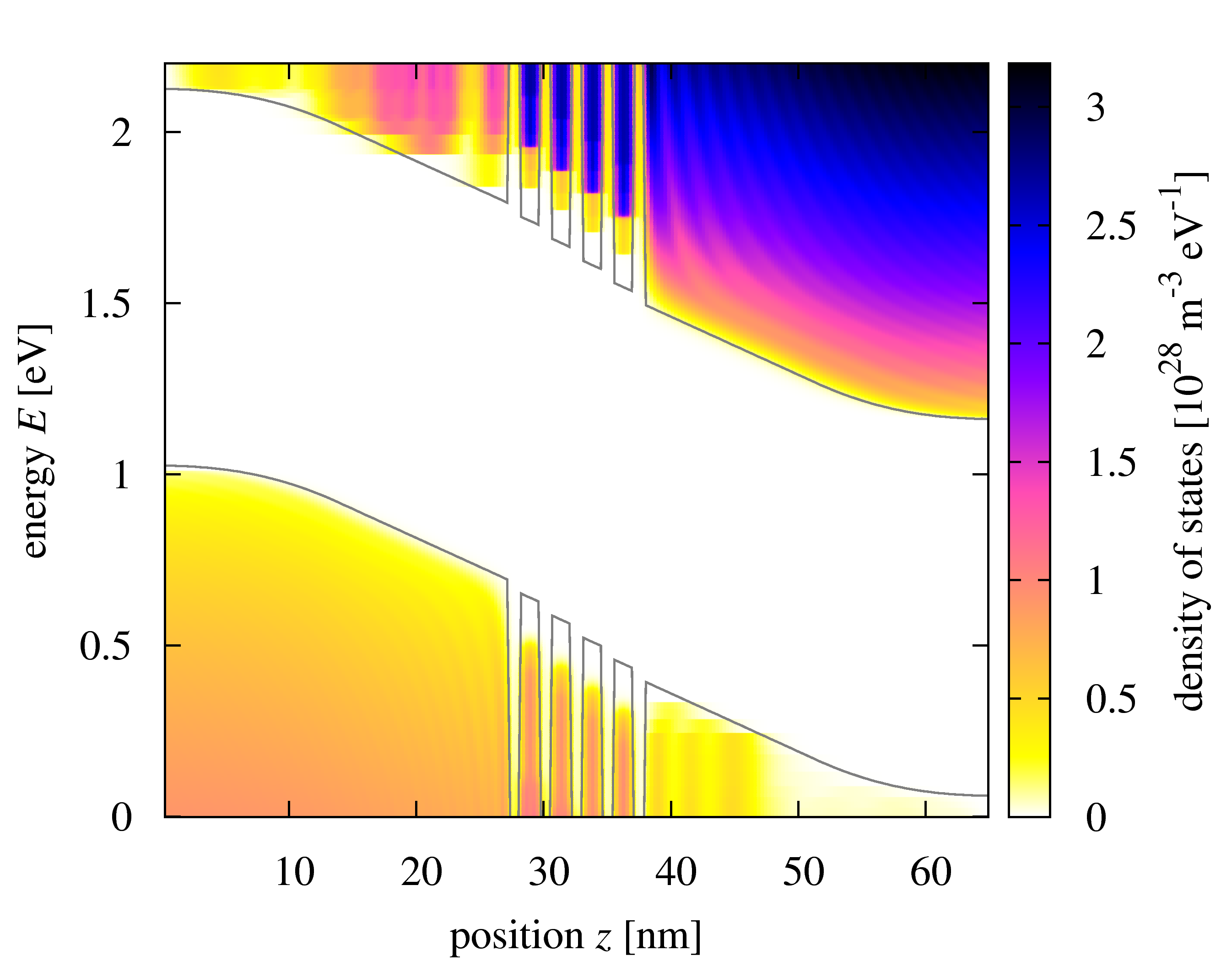}
\caption{Transverse momentum integrated local density of states of the $p$-$i(SL)$-$n$ photodiode at short circuit
conditions.\label{fig:ldos_int}}
\end{center}  
\end{figure}
Insertion of the oxide barriers leads to an increase of the effective band gap in the central region of the diode from 1.1 eV to 
$\sim$ 1.3 eV, as seen in Fig. \ref{fig:ldos_int}, which shows the transverse momentum integrated local density of states. 
In the actual situation of strong band bending, quantization also occurs in the form of notch states in front of the 
barriers. The density of states at minority carrier contacts is additionally depleted due to the imposition of closed system 
boundary conditions that prevents the formation of a dark leakage current under bias.
 \begin{figure}[b]
 \begin{center}
 \includegraphics[height=6cm]{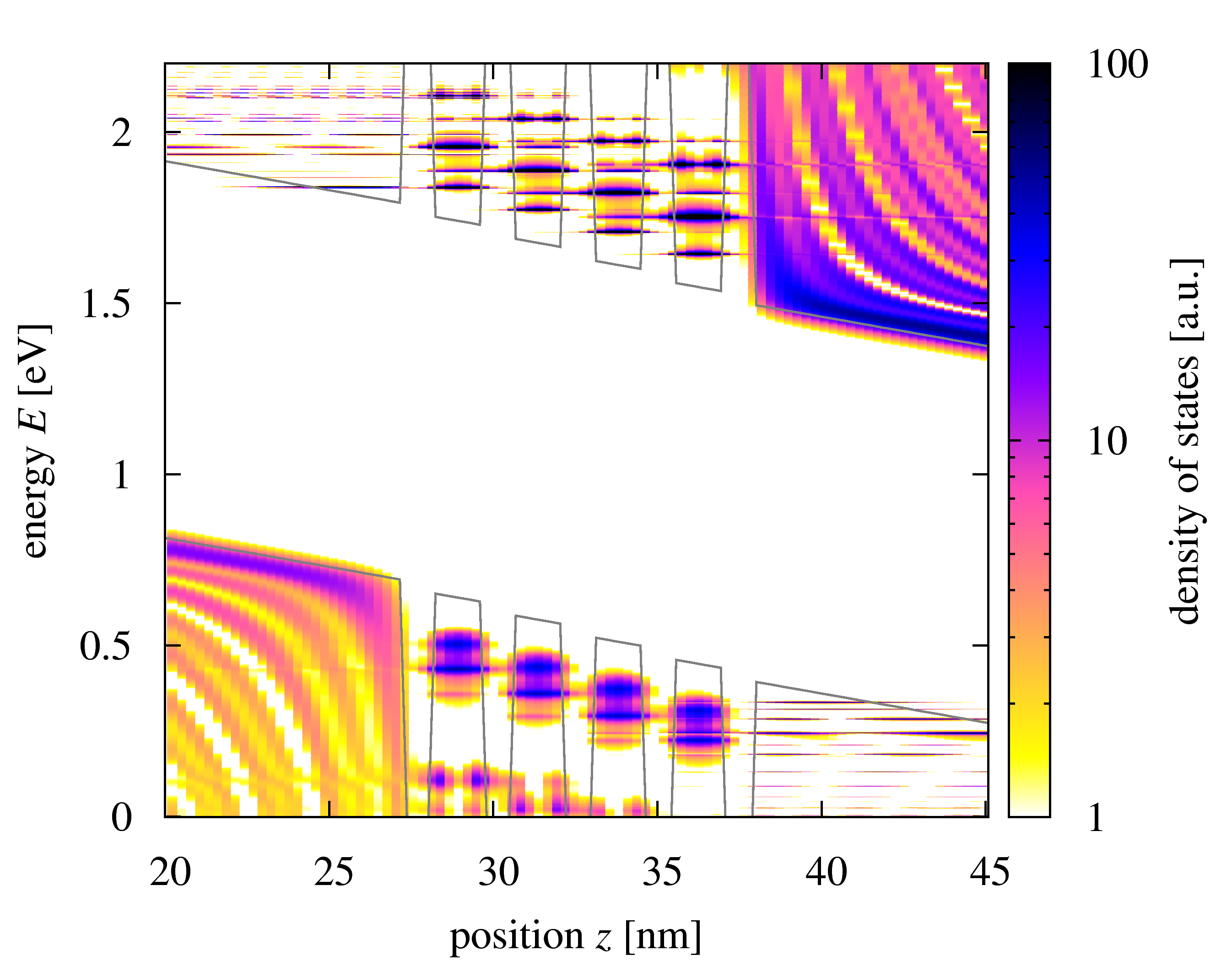}
\caption{Local density of states in the quantum well region at zero transverse momentum
($\mathbf{k}_{\parallel}=0$).\label{fig:ldos_k0}}
\end{center}
\end{figure}
The density of states component at zero transverse momentum displayed in Fig. \ref{fig:ldos_k0} allows the 
identification of the confined states in the different quantum wells, which are considerably localized due to the 
large internal field, with however finite overlap between neighboring wells in the case of the higher states. 
The ground state is split due to the different effective masses of the charge carriers, the effect being more 
pronounced for the electrons. 

\subsection{Generation and photocurrent spectrum}
\begin{figure*}[t!]
\begin{minipage}{8cm}
\begin{center}
 \includegraphics[height=6cm]{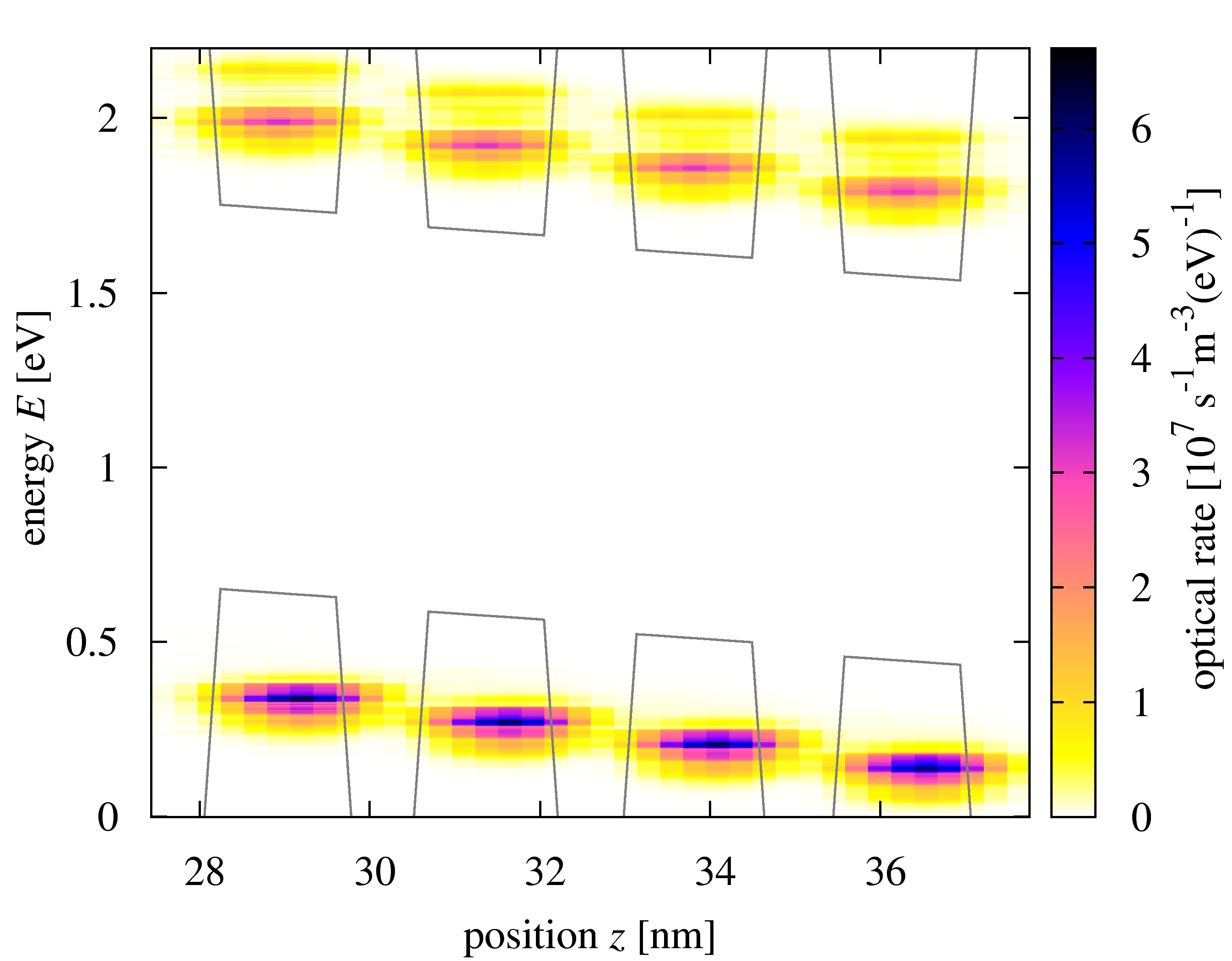}
 \end{center}
\caption{Spatially and energy resolved photon carrier photogeneration rate in the quantum well region at short circuit
conditions and under monochromatic illumination with energy $E_{\gamma}=1.65$ eV and intensity $I_{\gamma}=10$ kW/m$^{2}$.
\label{fig:photscattrate}}
\end{minipage}\qquad
\begin{minipage}{8cm}
\begin{center} 
 \includegraphics[height=6cm]{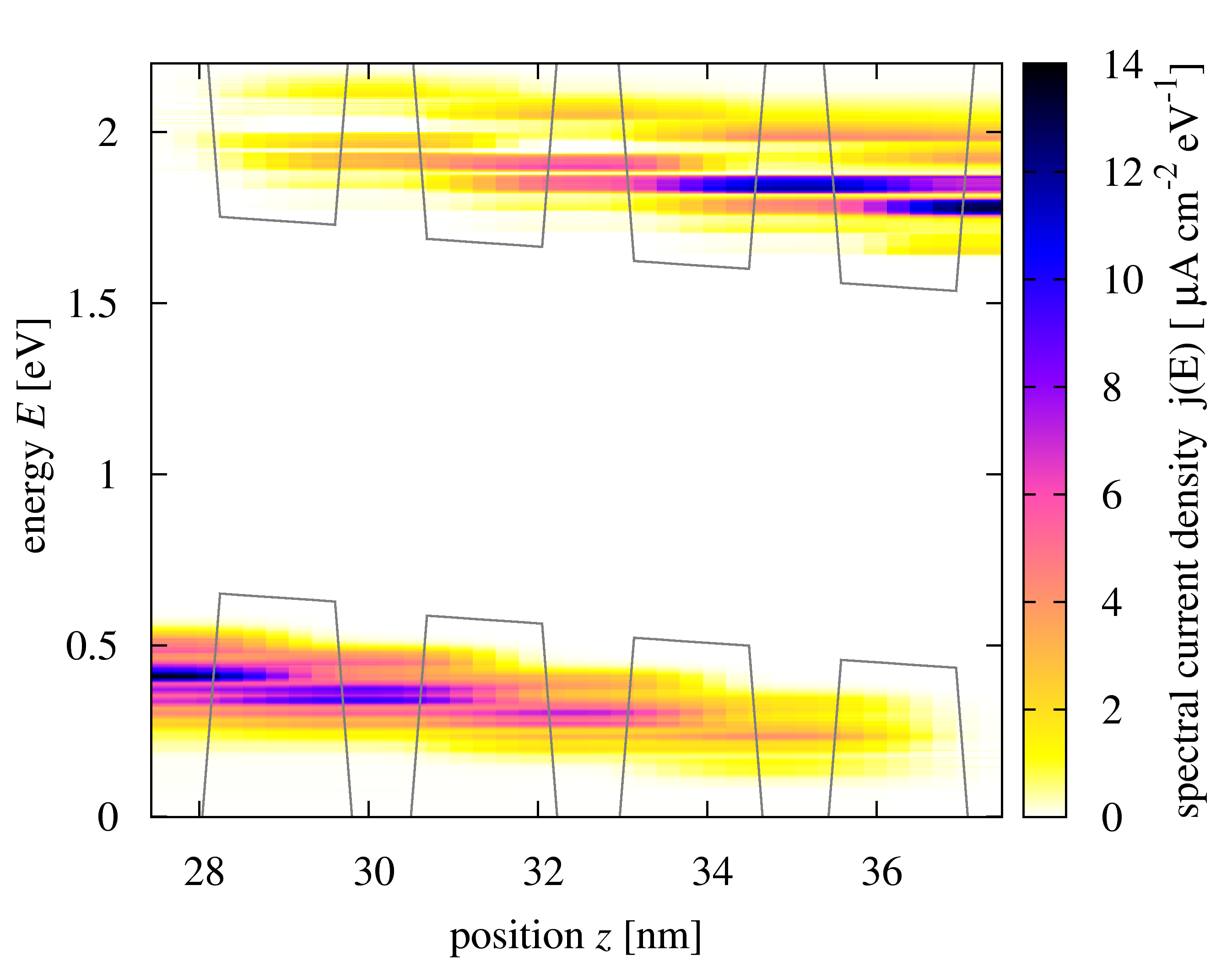}
 \end{center}
\caption{Spatially and energy resolved photon carrier short-circuit photocurrent density in the quantum well region
under monochromatic illumination with energy $E_{\gamma}=1.65$ eV and intensity $I_{\gamma}=10$ kW/m$^{2}$.
\label{fig:spectcurrdens}}
\end{minipage}
\end{figure*}

The spectral rate of carrier generation in the confined states under illumination with 
monochromatic light at photon energy $E_{\gamma}=1.65$ eV and intensity $I_{\gamma}=10$ kW/m$^{2}$ is shown in Fig. \ref{fig:photscattrate}. 
At this photon energy, both lowest and second minibands are populated. The photocurrent originating in this excitation
is shown in Fig. \ref{fig:spectcurrdens}. Current flows also in both first and second minibands, i.e. over the whole 
spectral range of generation, which means that relaxation due to scattering is not fast enough to confine transport to 
the band edge. However, transport of photocarriers is strongly affected by the inelastic interactions and is closest to the 
sequential tunneling regime. 

\section{\label{sec:conclusion}Conclusions}
In this paper, an adequate theoretical description of photogeneration and transport in Si-SiO$_{x}$ superlattice 
absorbers was presented. Based on quantum kinetic theory, the formalism allows a unified approach to both quantum 
optics and inelastic quantum transport and is thus able to capture pivotal features of photogeneration and photocarrier 
extraction in Si-based coupled quantum well structures, such as phonon-assisted optical transitions and field-dependent 
transport in superlattice states. Due to the microscopic nature of the theory, energy resolved information can be obtained, 
such as the spectra for photogeneration rate and photocurrent density, which shows that in the case of high internal fields, excess charge is
transported via sequential tunneling in the miniband where it is generated.

\section*{Acknowledgements}
Financial support was provided by the German Federal Ministry
of Education and Research (BMBF) under Grant No. 03SF0352E.

\bibliographystyle{spphys}  
   
\bibliography{/home/aeberurs/Biblio/bib_files/negf,/home/aeberurs/Biblio/bib_files/aeberurs,/home/aeberurs/Biblio/bib_files/generation,/home/aeberurs/Biblio/bib_files/scqmoptics,/home/aeberurs/Biblio/bib_files/pv,/home/aeberurs/Biblio/bib_files/sinova,/home/aeberurs/Biblio/bib_files/semiconductor_physics.bib}

\end{document}